\documentclass[
twocolumn,preprintnumbers,a4paper,superscriptaddress,aps,showpacs]{revtex4}
\usepackage{graphicx} 
\begin{document}

\def\la{\langle}
\def\ra{\rangle}
\def\om{\omega}
\def\Om{\Omega}
\def\vep{\varepsilon}
\def\wh{\widehat}
\def\P0{\wh{\cal P}_0}
\def\dt{\delta t}
\newcommand{\beq}{\begin{equation}}
\newcommand{\eeq}{\end{equation}}
\newcommand{\beqa}{\begin{eqnarray}}
\newcommand{\eeqa}{\end{eqnarray}}
\newcommand{\intf}{\int_{-\infty}^\infty}
\newcommand{\into}{\int_0^\infty}
\title{Transient response of a quantum wave to an instantaneous  
potential step switching}
\author{F. Delgado}
\affiliation{Departamento de Qu\'\i{}mica-F\'\i{}sica, Universidad del
Pa\'\i{}s Vasco, Apdo. 644, 48080 Bilbao, Spain}
\author{H. Cruz}
\affiliation{Departamento de F\'\i sica B\'asica, Universidad 
de La Laguna, Spain}
\author{J. G. Muga}
\affiliation{Departamento de Qu\'\i{}mica-F\'\i{}sica, Universidad del
Pa\'\i{}s Vasco, Apdo. 644, 48080 Bilbao, Spain}

\begin{abstract}
The transient response of a stationary  
state of a quantum particle in a step potential
to an instantaneous change in the step 
height (a simplified model for a sudden bias switch in 
an electronic  semiconductor device)
is solved exactly by means of a semianalytical 
expression. The characteristic times 
for the transient process up to the new stationary state
are identified. A comparison is made between the exact results and 
an approximate method. 
\end{abstract}
\maketitle
\section{Introduction}
Assume that due to a constant flux of incident quantum particles 
a stationary scattering state is formed
for a given one dimensional potential profile, and
that the asymptotic potential level is changed suddenly.
A physical realization would be an 
abrupt change in the bias voltage of an electronic device
\cite{MH88,Frensley90}. 
The wave will then respond to the 
potential switch until a new stationary state is attained 
for any finite position $x$.
Obtaining the characteristic time(s) of the transient 
is clearly of practical interest to determine
the transport properties
of small mesoscopic structures, but modelling the 
process by means of a grid discretization of
space in a ``finite box'' is far from simple
\cite{Frensley90}.
The problem is that the
boundary conditions at the box edges are not known a priori and involve 
simultaneous injection to and absorption from the
simulation (box) region.  
Some approximate ways to deal with the transients have been proposed
\cite{MH88,Frensley90,RRH91,YE95} 
but, surprisingly, no exact solution has been 
obtained up to now. Our aim in this paper is to work out an explicit
and exact solution of the transition between stationary monochromatic
waves due to an abrupt potential switch.
While the calculation is performed for simplicity for
a step potential that changes
the step height suddenly, other potential profiles, e.g., containing
square single or double barriers could be treated similarly.
Our results are in fact applicable to the outer regions 
of an arbitrary cut-off potential with different asymptotic levels
by inserting the appropriate reflection and transmission amplitudes. 
The basic trick to find the exact solution is to implement 
the action of the evolution operator of the 
new Hamiltonian on the initial state
using an integral expression in the complex momentum plane obtained by
Hammer, Weber and Zidell \cite{HWZ77}.

\section{Obtention of the exact expression}
For a potential step of the form 
\beq
V(x)=-V_0\Theta(x),
\eeq
where $\Theta(x)$ is the step function and $V_0>0$,   
a stationary state incident from the left has the form
\beq\label{initi}
\psi_0(x)=h^{-1/2}\left\{
\begin{array}{c}
e^{iq_0x/\hbar}+R_0^l(q_0)e^{-iq_0x/\hbar},\; x<0
\\
T_0^l(q_0) e^{ip_0x/\hbar},\;x\ge 0 
\end{array}
\right.
\eeq
where $p_0=(q_0^2+2mV_0)^{1/2}$ and $q_0$ are positive, and 
the reflection and transmission coefficients for left incidence 
are given by 
\beqa
R_0^l(q_0)=\frac{q_0-p_0}{q_0+p_0},
\\
T_0^l(q_0)=\frac{2q_0}{q_0+p_0}.
\eeqa
If the potential changes suddenly to 
\beq
\label{newv}
V(x)=-V_0'\Theta(x)
\eeq
at time $t=0$ the wave function 
will evolve in time. 
Finding $\psi(x,t)$ is equivalent to solve separately 
the time evolution of the initial functions
\beqa
\label{psi1}
\psi_1(x,t=0)&=&h^{-1/2}e^{iq_0x/\hbar}\Theta(-x),
\\  
\psi_2(x,t=0)&=&h^{-1/2}e^{-iq_0x/\hbar}\Theta(-x),
\\
\psi_3(x,t=0)&=&h^{-1/2}e^{ip_0x/\hbar}\Theta(x), 
\eeqa
and combine linearly the results with the appropriate coefficients:  
$\psi(x,t)=\psi_1(x,t)+R_0^{l}\psi_2(x,t)+
T_0^{l}\psi_3(x,t)$.
  
Of course the initial state could be different, in particular a state 
incident from the right. Clearly, to treat any possible 
initial stationary state we have to consider also a fourth truncated 
plane wave: 
\beq
\label{psi4}
\psi_4(x,t=0)=h^{-1/2}e^{-ip_0x/\hbar}\Theta(x).
\eeq
Moreover, we should also allow for the possibility of a purely 
imaginary $q_0$ in $\psi_2$ to describe the evolution of an initially 
evanescent wave. We shall now examine the four possible cases. Note that  
the evolution of each of these initial states is a realization 
of Moshinsky's shutter problem \cite{Moshinsky52} for the step potential. 
\subsection{$\psi_1$: initially a positive-momentum cutoff plane
wave in $x<0$.}
The momentum representation of $\psi_1(x,t=0)$ 
is given by 
\beq
\phi_1(q,t=0)=\frac{i}{2\pi}\frac{1}{q-q_0+i0}.
\eeq
Following \cite{HWZ77}, 
we shall rewrite the eigenstates of the new Hamiltonian,
\beq
\psi_q=h^{-1/2}\left\{
\begin{array}{l}
e^{iqx/\hbar}+R^l(q)e^{-iqx/\hbar},\;x\le 0
\\
e^{ipx/\hbar}T^l(q),\;x\ge 0
\end{array}
\right.,
\label{psiq}
\eeq
in the form
\beq
\label{r1}
\psi_q(x)=h^{-1/2}e^{iqx/\hbar}+{\cal R}_1,
\eeq
where 
\beq
\label{r1b}
{\cal R}_1=h^{-1/2}\left\{
\begin{array}{l}
R^le^{-iqx/\hbar}, \;x\le 0
\\
T^le^{ipx/\hbar}-e^{iqx/\hbar},\;x\ge 0
\end{array}
\right.
\eeq
and 
\beqa
E_q&=&q^2/2m=p^2/2m-V_0',
\\
q&=&[p^2-2mV_0']^{1/2},
\\
p&=&[q^2+2mV_0']^{1/2}.
\eeqa
The zero of energy is set by convention at the left level, and  
the amplitudes $R^l$ and $T^l$ 
are given in Appendix A. The square root in the definition of $q$ is 
chosen with a branch cut in the $p$-plane 
between the branch points $p=\pm(2mV_0')^{1/2}$, 
whereas $p$ has a branch cut in the $q$-plane between
$q=\pm i(2mV_0')^{1/2}$.
This means in particular that $q$ and $p$ have the same sign 
for $E_q>0$ and $-V_0'<E_q$. 
The solution method is based on writting the initial state as 
\beq
\psi_1(x,t=0)=\int_{C_{1}} dq\,\psi_q(x)\phi_1(q,t=0),
\eeq
where $C_{1}$ goes from $-\infty$ to 
$+\infty$ above all singularities (branch cut and pole). 
This is possible because 
\beq
\int_{C_{1}} dq\,\phi_1(q,t=0){\cal R}_1=0,
\eeq
as can be seen by closing the integration contour with a larg arc 
in the upper $q$-plane and using Cauchy's theorem.  
Since $\psi_q$ is an eigenstate of the Hamiltonian (even 
for complex $q$), the time 
dependent wavefunction is given by     
\beq
\label{psi1+}
\psi_1(x,t)=\int_{C_{1}} dq\,\psi_q(x)\phi(q,t=0)e^{-iE_qt/\hbar}. 
\eeq
\subsection{$\psi_2$: initially a negative-momentum cutoff plane wave 
in $x<0$, or an evanescent wave.}
The momentum representation of $\psi_2(x,t=0)$ is given by 
\beq
\phi_2(q, t=0)=\frac{i}{2\pi}\frac{1}{q+q_0+i0},
\eeq
with the pole again in the lower half $q$-plane. 
Following the same procedure used for $\psi_1$ and using the same 
${\cal R}$-function decomposition [Eqs. (\ref{r1}) and (\ref{r1b})],
the eigenfunctions $\psi_q$, and the same contour contour, 
$C_2=C_1$, one obtains 
\beq
\label{psi2+}
\psi_2(x,t)=\int_{C_2} dq\,\psi_q(x)\phi_2(q, t=0) e^{-iE_qt/\hbar}.
\eeq
In the evanescent case, $E_q<0$ and $q_0=i(2mV_0-p_0^2)^{1/2}$, 
the pole lies in the lower imaginary axis.  

\subsection{$\psi_3$: initially a positive momentum cutoff plane wave
in $x>0$.}
The treatment of $\psi_3(x)$ is different. 
The momentum representation of $\psi_3(x,t=0)$ is given by 
\beq
\phi_3(p, t=0)=\frac{-i}{2\pi}\frac{1}{p-p_0-i0}, 
\eeq 
with a pole in the upper half-plane.  
We shall use the eigenstates
\beq
\psi_p=h^{-1/2}\left\{
\begin{array}{l}
T^r(-p)e^{iqx/\hbar},\;x\le 0
\\
e^{ipx/\hbar}+R^r(-p)e^{-ipx/\hbar},\;x\ge 0
\end{array}
\right.
\label{psip}
\eeq
where the amplitudes are given in Appendix A. 
Similarly to Eq. (\ref{r1}) we write 
\beq
\psi_p(x)=h^{-1/2}e^{ipx/\hbar}+{\cal R}_3,
\eeq
with 
\beq
{\cal R}_3=
h^{-1/2}\left\{
\begin{array}{l}
T^r(-p)e^{iqx/\hbar}-e^{ipx/\hbar},\;x\le 0
\\
R^r(-p)e^{-ipx/\hbar},\;x\ge 0
\end{array}
\right.
\eeq 
The integral 
\beq
\int_{C_3}dp\, \phi_3(p,t=0){\cal R}_3=0
\eeq
vanishes for $C_3$ going from $-\infty$ to $\infty$ passing 
{\it below} the singularities (pole and branch cut). Note that the contour 
must now be closed in the lower half plane to apply Cauchy's theorem. 
Finally, 
\beq
\psi_3(x,t)=\int_{C_3}dp\,\phi_3(p,t=0)\psi_p(x)e^{-iE_qt/\hbar}.   
\eeq
\subsection{$\psi_4$: initially a negative momentum cutoff plane wave 
in $x>0$.}
The treatment is essentially the same as for $\psi_3$, but with   
\beq
\phi_4(p, t=0)=\frac{-i}{2\pi}\frac{1}{p+p_0-i0}. 
\eeq 
\section{Contour deformations}
The explicit expressions for $\psi_j(x,t)$, $j=1,..,4$ are 
given in Appendix B. The different terms  
contain exponentials of the form $\exp(\pm iqx/\hbar)$
or $\exp(\pm ipx/\hbar)$   
(for terms with support $x\le 0$ 
or $x\ge 0$ respectively).
In each case the integral 
is better solved in the corresponding plane, $q$ or $p$,  
by contour deformation along the steepest descent path
to be described below. We shall generically use the variable $k$ 
in both cases. 
Note that due to the decomposition of the stationary wave functions 
into three terms (associated with a $T$-amplitude,  
an $R$-amplitude and an independent term, $I$, see Eqs. (\ref{psiq}) and (\ref{psip}))
each wave function $\psi_j(x,t)$ may be separated into three
contributions that we shall denote as $\psi_{j\alpha}$,  
where $\alpha=I,T,R$:  
$\psi_j=\sum_\alpha \psi_{j,\alpha}$.   
There are twelve of these terms, each with support in one half-line, 
and therefore twelve different integrals.    
We shall denote as $k_j$ the poles in the  
momentum representation of the initial state $\psi_j(x,t=0)$. 
They are listed in Table 1 together with many other features 
of the twelve terms.    
%
%
%
%
The twelve terms may be written in the compact form 
\beqa
\psi_{j\alpha}(x)&=&c_jF{\cal I}_{j\alpha}
\\
{\cal I}_{j\alpha}&=&\int_{C_j} dk\, e^{-i(ak^2+kb_{j\alpha})}g_{j\alpha}(k),
\eeqa
and vanish outside their support region.   
In the above expressions $c_{1,2}=i/2\pi h^{1/2}$, $c_{3,4}=-i/2\pi h^{1/2}$, 
$a=t/(2m\hbar)$, 
\beq
F=\Theta(x)(e^{itV_0'/\hbar}-1)+1,
\eeq
and  
$g(k)$ has always a pole (we shall drop the subscripts $j,\alpha$ unless
they are strictly necessary). Moreover, the $R$ and $T$-terms 
have also a branch cut
singularity. The saddle point of the exponent is at $k=-b/2a$
($xm/t$ for $I$ and $T$-terms and $-xm/t$ for $R$-terms) 
and the steepest descent path is the straight line
${\rm Im}(k)=-({\rm Re}(k)+b/2a)$.   
By completing the square, introducing the new variable $u$, 
\beq
u=(k+b/2a)/f,\;\;\;\; f=(1-i)(m\hbar/t)^{1/2}, 
\eeq
which is real on the steepest descent path and zero at the 
saddle point, and mapping the contour to the $u$-plane, 
the integral takes the form
\beq
{\cal I}=e^{imx^2/\hbar t}f\int_{C_u} du\,e^{-u^2} G(u),
\eeq
where $G(u)\equiv g[k(u)]$. This function has a simple pole at 
$u_0\equiv (k_0+b/2a)/f$ and possibly a branch cut,  
whereas  $k_0$ is given in Table 1.   
It is now useful to separate the pole and branch-cut contributions
explicitly and 
write $G$ as 
\beq
G(u)=\frac{A_0/f}{u-u_0}+H(u), 
\eeq
where $A_0/f$ is the residue of $G(u)$ at $u=u_0$ and the remainder, $H(u)$,
is obtained by substraction. Note that $H(u)$ is either an 
entire function, if there is no branch cut, 
or its only singularity is the branch cut.    
 
\begin{figure}
{\includegraphics[angle=-90,width=3in]{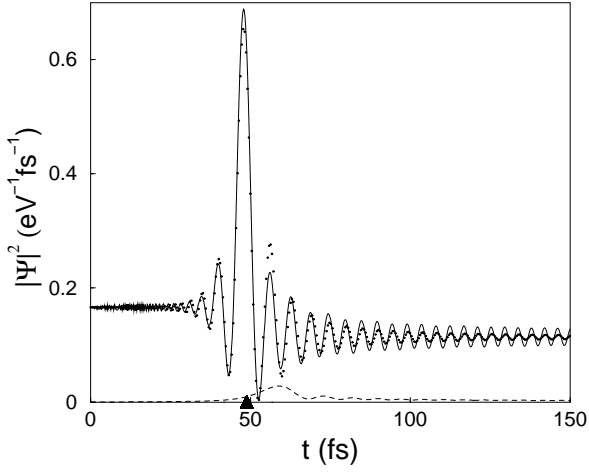}}
\caption{Exact density versus $t$ (solid line), and approximation 
using Eq. (\ref{aapro}) (dotted line) for $x=100$nm. Also shown is the 
contribution of ${\cal I}''$ multiplied by a factor of 10 (dashed line), and the 
point when the steepest descent path crosses the branch point (triangle). 
Mass=0.067$m_e$, 
$E_q=0.3$eV, 
$V_0=0.3$eV, 
$V_0'=0.8$eV.
}
\label{fapr}
\end{figure}

\begin{figure}
{\includegraphics[angle=-90,width=3in]{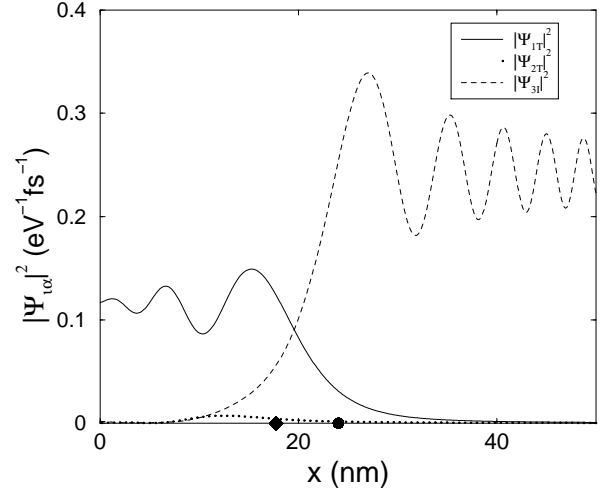}}
\caption{$|\psi_{1T}|^2$, $|\psi_{2T}|^2$ and $|\psi_{3I}|^2$ 
at $t=10$ fs versus $x$. mass=0.067$m_e$, 
$E_q=0.3$eV, 
$V_0=0.3$eV, 
$V_0'=0.8$eV.
}
\label{fotoi3}
\end{figure}

The integral ${\cal I}$ is thus separated into two integrals, 
${\cal I}={\cal I}'+{\cal I}''$.   
The first one may be reduced to a known function by deforming the contour 
along the steepest descent 
path (real-$u$ axis) and taking proper care of the pole contribution,      
\beqa
{\cal I}'&\equiv& e^{imx^2/\hbar t} A_0\int_{C_u} du\,\frac{e^{-u^2}}{u-u_0}
\nonumber\\
&=&\left\{
\begin{array}{ll}
-i\pi e^{imx^2/\hbar t} A_0 w(-u_0), &j=1,2
\\
i\pi e^{imx^2/\hbar t} A_0 w(u_0), &j=3,4
\end{array}
\right.
\eeqa
where $w(z)=\exp(-z^2) {\rm erfc}(-iz)$. 
In general the second integral, which involves the remainder $H$, 
has to be evaluated numerically.
\beq
{\cal I}''\equiv e^{imx^2/\hbar t}f \int_{C_u} du\,H(u).
\eeq
However, the computational effort is greatly reduced by deforming
the contour along the steepest descent path too.
The branch cut, whenever it is present, cannot be crossed and 
has to be surrounded. 
This occurs for  
\beq
\label{cond}
|mx/t|<(2mV_0')^{1/2}. 
\eeq
Otherwise there is no branch cut contribution and the integral may
be expressed as a series by expanding $H(u)$ around the 
origin and integrating term by term,        
\beqa
{\cal I}''&=&e^{imx^2/\hbar t }f\pi^{1/2}\Bigg[ H(u=0)
\\
&+&
\left.\sum_{n=1}^\infty \frac{1\times3\times...\times(2n-1)}
{2^n(2n)\!}H^{(2n)}(u=0)\right],
\nonumber
\eeqa
In practice the first term gives already a very good
approximation, even if Eq. (\ref{cond}) holds.
The ${\cal I}''$-integrals 
have to be calculated numerically only for $|mx/t|<(2mV'_0)^{1/2}$
and their relative 
importance with respect to $w$-terms from ${\cal I}'$ is only
significant for rather small $x$ at intermediate times, since as
$t\to\infty$, ${\cal I}''\to 0$ in all cases.     
A general analytical approximation making use of 
${\cal I}'$ and the first term in ${\cal I}''$ is  given by    
\beq
\psi_{j\alpha}\approx 
\frac{Fe^{imx^2/2\hbar t}}{2h^{1/2}}\left[\pm A_0 w(\mp u_0)+ 
\frac{f}{\pi^{1/2}} H(u=0)\right]
\label{aapro}
\eeq
with the upper sign for $j=1,2$ and the lower sign for $j=3,4$, see Fig. \ref{fapr}. 
Note the basic role of the $w$-functions,
which may be considered elementary transient mode
propagators   
of the Schr\"odinger equation \cite{Nussenzveig92}.
They will show approximate wave fronts 
when $x(u_0=0)$ lies within the domain of the term
(i.e., when the saddle meets the pole).
This occurs (for $q_0>0$) 
for the terms 1T, 1R, 2I, 3I, 4T and 4R, see two examples
in Figure \ref{fotoi3}.  

The long time behaviour may be obtained from 
the asymptotic (large-$z$) formula, 
\beq
w(z)\sim\left\{
\begin{array}{ll}
\frac{i}{\pi^{1/2}z}, &{\rm Im}\,z>0,
\\
\frac{i}{\pi^{1/2}z}+2e^{-z^2},&{\rm Im}\,z<0.
\end{array}
\right.
\label{asy}
\eeq
All $\psi_{2,3}$ terms vanish as $t\to\infty$ for finite $x$,
since these waves move initially away from the  origin. 
In spite of this dominant motion, note that there is a transitory 
and generally small contribution of $\psi_{2T}$ at positive $x$ and 
of $\psi_{3T}$ at negative $x$. 
On the contrary the $w$-functions of $\psi_{1,4}$ pick up the exponential 
contribution in Eq. (\ref{asy}) which gives the new stationary states.   
In particular, as $t\to\infty$ and for finite $x$,  
\beqa
\nonumber
\psi_1(x,t)&\to& \frac{e^{-iE_qt/\hbar}}{h^{1/2}}\left\{
\begin{array}{ll}
e^{iq_0x/\hbar}+R^l(q_0) e^{-iq_0x/\hbar},&x<0
\\
T^l(q_0) e^{ip_0'x/\hbar},&x\ge 0 
\end{array}
\right.
\\
\psi_4(x,t)&\to& \frac{e^{-iE_qt/\hbar}}{h^{1/2}}\left\{
\begin{array}{ll}
T^r(p_0')e^{-iq_0x/\hbar},&x<0
\\
e^{-ip_0'x/\hbar}+R^r(p_0')e^{ip_0'x/\hbar},&x\ge 0 
\end{array}
\right.
\nonumber
\eeqa

\section{Examples}
\begin{figure}
{\includegraphics[angle=-90,width=3in]{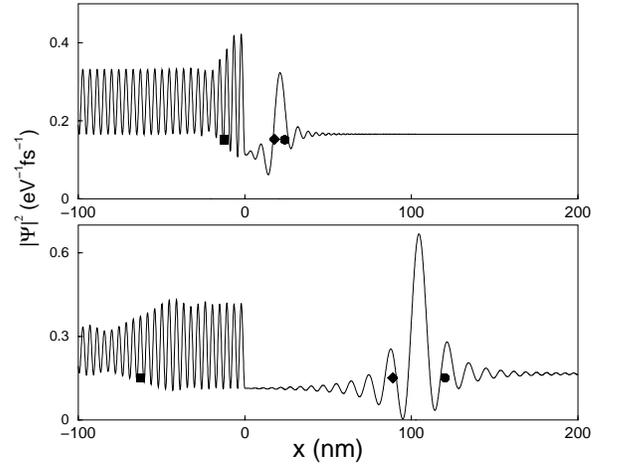}}
\caption{$|\psi|^2$ versus $x$ for $t=10$ fs (upper figure) and $t=50$ fs
(lower figure). 
Exact solution (solid line).  
The diamond marks $p_0t/m$; the circle $p_0't/m$; the square $-q_0t/m$.
mass=0.067$m_e$, 
$E_q=0.3$eV, 
$V_0=0.3$eV, 
$V_0'=0.8$eV.
}
\label{fotoM}
\end{figure}
\begin{figure}
{\includegraphics[
angle=-90,width=3in]{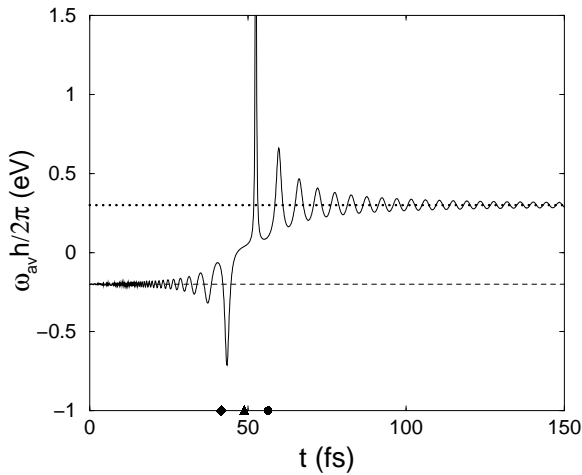}}
\caption{$\hbar\times\omega_{av}$ versus time at $x=100$nm,
where the average local frequency
is defined as $\omega_{av}\equiv-{\rm Im}[(d\psi/dt)/\psi]$. Mass=0.067$m_e$, 
$E_q=0.3$eV, 
$V_0=0.3$eV, 
$V_0'=0.8$eV. The diamond and circle indicate $xm/p_0'$ and $xm/p_0$ respectively. 
The initial value is $V_0+E_q-V_0'$ whereas the final asymptotic value is $E_q$.}
\label{wp3}
\end{figure}

Figure \ref{fotoM} shows a typical wavefunction ``density''
\footnote{Because of the 
scattering normalization of the wavefunctions the ``densities'' do not have 
dimensions [1/L], which would of course be obtained by forming wave packets.}
versus $x$ at two  
fixed instants $t_2>t_1>0$ for $V_0'>V_0>0$. For $x>0$
the main features are two flatter  regions
representing the old (to the right) and new (to the left) stationary
regimes separated by an oscillating structure. A simple semiclassical 
picture provides a good zeroth order explanation: assume a 
stationary flux of classical particles in the old potential, with 
incident momentum $q_0$ and transmitted momentum $p_0$. After the potential 
switch at $t=0$, the last transmitted particle with momentum 
$p_0$ will be at $p_0t/m$, whereas the first transmitted particle 
with momentum $p_0'$ will be at $p_0't/m$.  Since $p_0'>p_0$ 
there is a region of width $(p_0'-p_0)t/m$ where the two types of
particles coexist. In the corresponding quantum scenario one may expect 
interference and oscillations of
wavelenght $2\pi\hbar/(p_0'-p_0)$ in this region, whereas the regions dominated by only 
one plane wave the density does not oscillate and is proportional to the 
corresponding transmission probability, either $|T_0^l(q_0)|^2$ for the 
``old wave'' or $|T^l(q_0)|^2$ for the ``new wave'', which in the present case 
is smaller than the former, a somewhat surprising feature of quantum 
scattering off potential steps from the perspective of classical mechanics.
The two crytical positions
are marked in the figure with a diamond and a circle. 
The average local frequency \cite{Cohen95,MB00} shows also the transition between the two regimes, 
see Fig. \ref{wp3}.

The $x<0$ region is clearly divided into 
``old'' (to the left) and ``new'' regimes where incident and reflected 
components interfere. They admit also a simple
analysis: since the reflected wave stays dominated by 
momentum $-q_0$, the interference pattern wavelenght stays the same
in the new and old regimes, $(\pi\hbar/q_0)$,
and the only difference is the 
amplitude change due to the change of reflection probability
from $|R_0^l(q_0)|^2$ to $|R^l(q_0)|^2$. The transition at 
$x=-q_0 t/m$ is marked with a square in Fig. \ref{fotoM}.        
        
In the example shown the dominant terms  for $x>0$  are $\psi_{3T}$
and $\psi_{1I}$ representing respectively the ``old'' and ``new'' wave.
Eq. (\ref{aapro}) with these
two terms only provides a very 
good approximation. For $x<0$ the dominant terms are 
$\psi_{1I}$ (incident wave),  $\psi_{2I}$ (old reflected wave)
and $\psi_{1R}$ (new reflected wave). Again, the analytical approximation 
describes the main features correctly. 
We may expect a worse performance of the analytical approximation 
in processes involving tunnelling or evanescent waves, with the pole 
lying close to the branch cut: for example, 
when $E_q<0$, or $-V_0'>E_q>0$. Some of these
processes and the corresponding time scales have been studied recently 
in \cite{GVDM02} so we shall not insist on them here.

\begin{figure}
{\includegraphics[angle=-90,width=3in]{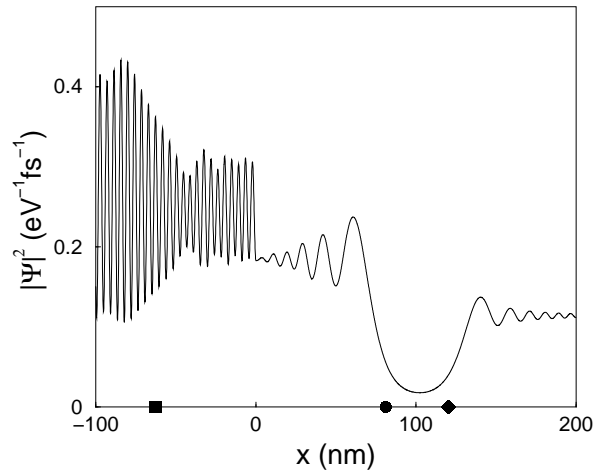}}
\caption{$|\psi|^2$ versus $x$:  
mass=0.067$m_e$, 
$E_q=0.3$eV, 
$V_0=0.8$eV, 
$V_0'=0.2$eV.
The square is at $-q_0t/m$, the circle at $p_0't/m$ and the diamond at 
$p_0t/m$.} 
\label{foto5}
\end{figure}

Fig. \ref{foto5} shows the density for a case in which $V_0>V_0'>0$. 
The new wave moves now at a slower pace than the old one so there is no 
interference structure between  
the two.

\section{Comparison with an approximate method}

\begin{figure}
{\includegraphics[angle=-90,width=3in]{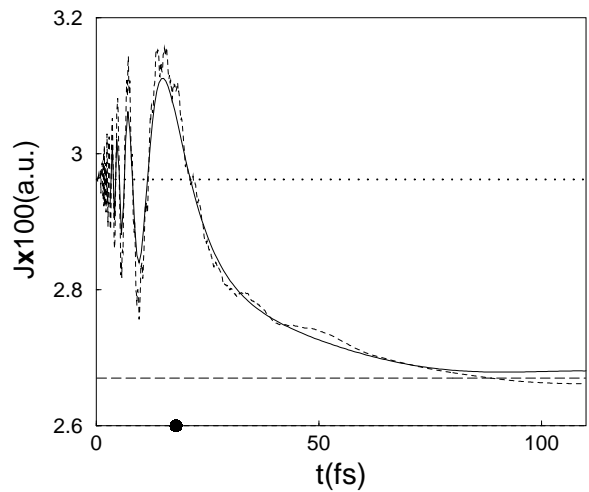}}
\caption{$J\times100$ at the left box edge,  $x=-22.48$nm.  mass=0.042$m_e$,
$E_q=0.04$eV, 
$V_0=0.42$eV, 
$V_0'=0.62$eV;    
box lenght: $L=44.96$nm; number of grid points: $N=10^4$; 
time step: $\Delta t=10 ^{-6}$fs.
The circle is at $t=xm/q_0$. The two stright lines indicate the 
values of $J\times100$ for the initial and final stationary functions. }
\label{j-l}
\end{figure}

\begin{figure}
{\includegraphics[angle=-90,width=3in]{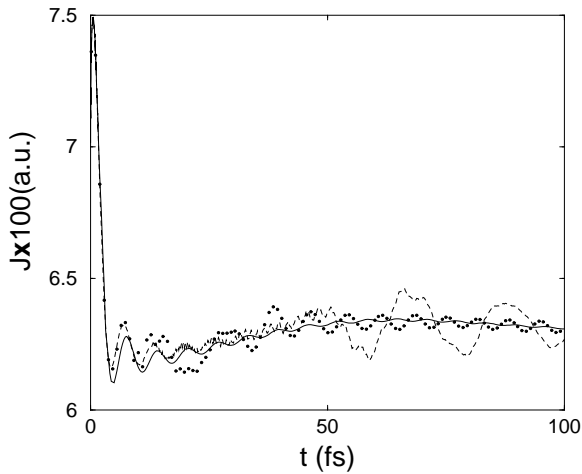}}
\caption{$J\times100$ at $x=0$: exact (solid line) and for two different 
box lenghts:   
$L=179.84$nm, $N$=80000, $\Delta t=5.e-5$fs (dashed line);    
and $L=44.96nm$ with the same grid  density and $\Delta t$
as in the other box (dots).  
mass=0.042$m_e$, 
$E_q=0.04$eV, 
$V_0=0.42$eV, 
$V_0'=0.62$eV
}
\label{j0c}
\end{figure}

\begin{figure}
{\includegraphics[angle=-90,width=3in]{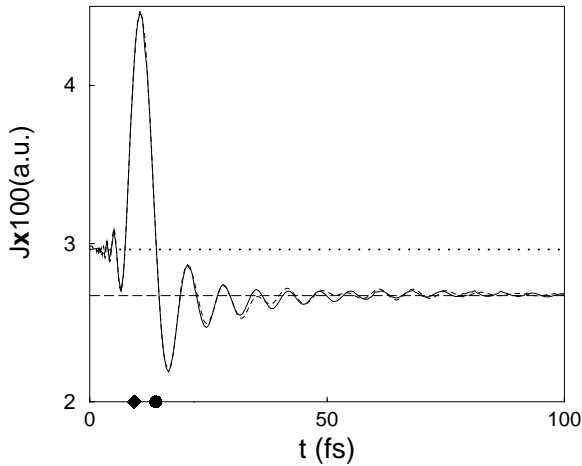}}
\caption{$J\times100$ at the right box edge,  $x=22.48$nm.  mass=0.042$m_e$,
$E_q=0.04$eV, 
$V_0=0.42$eV, 
$V_0'=0.62$eV;    
box lenght: $L=44.96$nm; number of grid points: $N=10^4$; 
time step: $\Delta t=10^{-6}$fs.   
The circle is at $xm/p_0'$ and the diamond at $xm/p_0$.
The two stright lines indicate the values for initial 
and final  
stationary functions.}
\label{jl}
\end{figure}

As stated in the Introduction, the basic difficulty to deal with  
transient phenomena between stationary scattering states 
by means of grid methods is that the time dependent boundary
conditions at the box edges are not known a priori. 
Several approximate schemes have been proposed to overcome this 
difficulty, and our exact solution provides a needed reference to test their 
validity and/or range of applicability. 
We have in particular made a comparison with a method  
proposed by Mains and Haddad \cite{MH88} based on looking 
a short distance into the simulation domain to determine what is coming 
out. Especifically, at the left edge region the wave is written as   
\beq
\psi_l=Ae^{iq_0x/\hbar}+B(x,t)e^{-iq_0x/\hbar},
\eeq
Substituting this form in the Schr\"odinger equation and neglecting 
the second order derivative of $B$,
\beq
i\hbar\frac{\partial\psi_l}{\partial t}\approx\frac{q_0^2}{2 m}\psi_l
+i\frac{\hbar q_0}{m}\frac{\partial B(x,t)}{\partial x}e^{-iq_0x/\hbar}.
\eeq
The first derivative of $B$ is then calculated numerically at each 
time step with the first two spatial points, and is used to update 
the boundary condition for the next time step as  
\beq
\psi(t+\Delta t)_l\approx \psi(t)_le^{-iE_{q_0}\Delta t/\hbar}
+\frac{q_0}{m}\frac{\partial B(x,t)}{\partial x} e^{-iq_0x/\hbar} \Delta t.
\eeq
Similarly, the wave function at the right edge grid points is written as 
\beq
\psi_r=C(x,t)e^{ip_0'x/\hbar}.
\eeq
Assuming again that $C$ is linear in $x$ and evaluating its derivative numerically 
with the two last grid points the boundary condition to the right is
updated as 
\beq
\psi(t+\Delta t)_r\approx \psi(t)_re^{-iE_{q_0}\Delta t/\hbar}
-\frac{p_0'}{m}\frac{\partial C(x,t)}{\partial x} e^{-ip_0'x/\hbar} \Delta t.
\eeq
We have adapted Koonin's grid method \cite{Koonin85} 
based on the Caley transform 
to this boundary-condition scheme and have calculated
the ``flux'' \footnote{The ``flux'' $J(x)$
is computed with the standard expression $\frac{\hbar}{m}{\rm Im}[\psi(x)^*d\psi(x)/dx]$. 
However, because of the continuum normalization of the wave functions $J$ does not have 
dimensions of a current density, which would be recovered by forming a
normalizable 
wave packet.}   
versus time at the box edges (Figures \ref{j-l} and \ref{jl})
and at the center $x=0$ (Figure \ref{j0c}) for the same 
potential jump considered in \cite{MH88}; 
the effective mass is taken as 
$m=0.042$ au (for In$_{0.53}$Ga$_{0.47}$As-AlAs) and 
the inicident energy corresponds to the Fermi level.   

The comparison with the exact results demonstrates that the linear ansatz
is quite good at the right edge but fails at the left edge, where 
incident and reflected components interfere. The error introduced 
at the left edge propagates and affects eventually to whole 
simulation domain, in particular the flux at the origin is deformed
with respect to the exact one quite rapidly. 
Enlarging the box retardates the deviation from the exact result, 
see Fig. \ref{j0c},  
but the computational cost becomes exceedingly large to 
reproduce correctly the whole transient at the origin.

\section{Discussion and conclusions}

Since its discovery by Tsu and Esaki, tunneling through semiconductor
nanostructures has been the object of a great attention due to its
possible applications to ultrahigh speed electronic devices \cite{h1}. With
the development of novel semiconductor nanostructures, it has become
important to carry out theoretical and experimental studies on the
tunneling process of carriers when an external bias is applied. In this way,
electric field-induced electron transport has been recently explored in
quantum dot
arrays \cite{h2}, resonant tunneling diodes \cite{h3}, and semiconductor
superlattices \cite{h4}. However, we note that one remaining key question
in these experiments is the analysis of the device transient response
to an instantaneous potential step
switching. The characteristic time of the response is of
practical interest to determine the nanostructure transport properties
and its possible applications to novel ultrahigh speed semiconductor
devices.

We have obtained an exact solution of the transition between two stationary 
scattering states due to the sudden change in a potential step. Equivalently, 
we have solved exactly the Moshinski shutter problem for an arbitrary 
cut-off plane wave in the step potential. (For other potential shapes  
see \cite{TKF87,JJ89,Kleber94,BM96,GCV01}). The explicit expressions obtained
for their time evolution would in fact be directly applicably to
an arbitrary cut-off potential
with different asymptotic levels by using the appropriate transmission 
and reflection amplitudes. (For recent work on 
step-like potentials scattering see \cite{BEM01a,BEM01b}).  
The exact results allow to identify characteristic times for
the transients. They also provide a needed reference for testing approximate 
methods that model time dependent open systems (finite
systems exchanging particles 
with the outside) with injecting and absorbing boundary conditions.

\begin{acknowledgments}
We are grateful to S. Brouard and I. L. Egusquiza for many useful discussions.  
JGM and FD acknowledge support 
by Ministerio de Ciencia y Tecnolog\'\i a (BFM2000-0816-C03-03), 
UPV-EHU (00039.310-13507/2001), and the Basque Government (PI-1999-28).
\end{acknowledgments}

\appendix
\section{Reflection and transmission amplitudes}
The reflection and transmission amplitudes 
in stationary waves with left and right incidence
are given by 
\beq
\begin{array}{ll}
R^l(q)=\frac{q-p}{q+p},&T^l(q)=\frac{2q}{q+p}
\\
R^r(p)=\frac{p-q}{q+p},&T^r(p)=\frac{2p}{q+p}
\end{array}
\eeq
for positive values of the arguments. 
The analytical continuations for negative arguments 
are the amplitudes for the ``outgoing'' or
``time-reversed'' stationary states. 
\section{Wave functions}
These are the time dependent wave functions corresponding to the initial conditions 
given in Eqs. (\ref{psi1}-\ref{psi4}) for the potential of Eq. 
(\ref{newv}):  
\beq
\psi_{1,2}=
\frac{i}{2\pi \sqrt{h}}\left\{
\begin{array}{l}
\int_{C_j} dq\, \frac{e^{-iE_qt/\hbar}}
{q\mp q_0+i0}(e^{iqx/\hbar}+\frac{q-p}{q+p}e^{-iqx/\hbar})
,\;x\le 0
\\
\int_{C_j} dp\,\frac{2p e^{ipx/\hbar} e^{-iE_qt/\hbar}}{(q+p)(q\mp q_0+i0)}
,\;x\ge 0
\end{array}
\right.
\eeq 
where the minus and plus signs correspond to $\psi_1$ and $\psi_2$
respectively, and 
\beq
\psi_{3,4}=
\frac{-i}{2\pi \sqrt{h}}\left\{
\begin{array}{l}
\int_{C_j} dq\, \frac{2q e^{iqx/\hbar}e^{-iE_qt/\hbar}}{(q+p)(p\mp p_0-i0)},\;x\le 0
\\
\int_{C_j} dp\,\frac{e^{-iE_qt/\hbar}}{p\mp p_0-i0}
(e^{ipx/\hbar}+\frac{p-q}{q+p}e^{-ipx/\hbar})
,\;x\ge 0
\end{array}
\right.
\eeq
with the minus sign for $\psi_3$ and the plus sign for $\psi_4$.

\begin{center}
\begingroup
\squeezetable
\begin{table*}
\begin{tabular}{|c|c|c|c|c|c|c|c|c|c|}
\hline
Term&support&-b/2a&pole&H-term&$e^{iV_0't/\hbar}$&
contour&$A_0$&$k_0$
&$g$
\\
\hline
$\psi_{1T}$&$x>0$&$xm/t>0$&$q_0-i0$&yes&yes&above&$T^l(q_0-i0)$&$p(q_0-i0)$
&$\frac{p}{q}\frac{T^l(p)}{q-q_0+i0}$
\\
\hline
$\psi_{1I}$&$x<0$&$xm/t<0$&$q_0-i0$&no&no&above&$1$&$q_0-i0$
&$\frac{1}{q-q_0+i0}$
\\
\hline
$\psi_{1R}$&$x<0$&$-xm/t>0$&$q_0-i0$&yes&no&above&$R^l(q_0-i0)$&$q_0-i0$
&$\frac{R^{l}(q)}{q-q_0+i0}$
\\
\hline
$\psi_{2T}$&$x>0$&$xm/t>0$&$-q_0-i0$&yes&yes&above&$T^l(-q_0-i0)$&$p(-q_0-i0)$
&$\frac{p}{q}\frac{T^l(p)}{q+q_0+i0}$
\\
\hline
$\psi_{2I}$&$x<0$&$xm/t<0$&$-q_0-i0$&no&no&above&$1$&$-q_0-i0$
&$\frac{1}{q+q_0+i0}$
\\
\hline
$\psi_{2R}$&$x<0$&$-xm/t>0$&$-q_0-i0$&yes&no&above&$R^l(-q_0-i0)$&$-q_0-i0$
&$\frac{R^{l}(q)}{q+q_0+i0}$
\\
\hline
$\psi_{3I}$&$x>0$&$xm/t>0$&$p_0+i0$&no&yes&below&$1$&$p_0+i0$
&$\frac{1}{p-p_0-i0}$
\\
\hline
$\psi_{3R}$&$x>0$&$-xm/t<0$&$p_0+i0$&yes&yes&below&$R^r(-p_0-i0)$&$p_0+i0$
&$\frac{R^r(-p)}{p-p_0-i0}$
\\
\hline
$\psi_{3T}$&$x<0$&$xm/t<0$&$p_0+i0$&yes&no&below&$T^r(-p_0-i0)$&$q(p_0+i0)$
&$\frac{q}{p}\frac{T^r(-p)}{p-p_0-i0}$
\\
\hline
$\psi_{4I}$&$x>0$&$xm/t>0$&$-p_0+i0$&no&yes&below&$1$&$-p_0+i0$
&$\frac{1}{p+p_0-i0}$
\\
\hline
$\psi_{4R}$&$x>0$&$-xm/t<0$&$-p_0+i0$&yes&yes&below&$R^r(p_0-i0)$&$-p_0+i0$
&$\frac{R^r(-p)}{p-p_0-i0}$
\\
\hline
$\psi_{4T}$&$x<0$&$xm/t<0$&$-p_0+i0$&yes&no&below&$T^r(p_0-i0)$&$q(-p_0+i0)$
&$\frac{q}{p}\frac{T^r(-p)}{p-p_0-i0}$
\\
\hline
\end{tabular}
\caption{Features of the terms $\psi_{j\alpha}$.}
\end{table*}
\endgroup
\end{center}

\end{document}